\documentclass[journal]{IEEEtran}
 \pdfoutput=1

\usepackage[absolute,overlay]{textpos}

\usepackage{balance}
\usepackage{graphicx}
\usepackage [noadjust]{cite} %[nospace]
\usepackage{bm}  % bold math for greek symbols
\usepackage{amsmath}
\usepackage{amssymb}
\usepackage{enumerate}
\usepackage{stfloats}
\usepackage{cases}
\usepackage{epstopdf}
\usepackage{soul,color} % \ul  \hl
\usepackage{hyperref}
\usepackage[percent]{overpic}
\usepackage{tabularx}
\usepackage[table]{xcolor}
\usepackage{multirow}
\usepackage{booktabs}  % multicolumn tabular
\usepackage{tabu} % different font sizes in different rows of tabular
\newcommand{\TB}{\boldsymbol{\Theta}}

\newcommand{\argmin}{\operatornamewithlimits{argmin}}

\begin{document}
	\raggedbottom
	\allowdisplaybreaks
    \title{Practical Challenges for Reliable RIS Deployment in Heterogeneous Multi-Operator Multi-Band Networks  
  
  \thanks{%This research is supported by Business Finland via project 6GBridge - Local 6G (Grant Number: 8002/31/2022), and Research Council of Finland, 6G Flagship Programme (Grant Number: 369116).% 346208).
  This research is supported by Research Council of Finland via 6G Flagship Programme under Grant 369116, Research Council of Finland via Profi6 Project under Grant 336449, Business Finland via  6GBridge - Local 6G Project under Grant 8002/31/2022, and Business Finland via REEVA Project under Grant 10284/31/2022.
    }
    \thanks{The work of M. Di Renzo was supported in part by the Nokia
Foundation, the French Institute of Finland, and the French Embassy in
Finland under the France-Nokia Chair of Excellence in ICT. %He was also supported in part by the European Commission through the Horizon Europe project COVER under grant agreement number 101086228, the Horizon Europe project UNITE under grant agreement number 101129618, and the Horizon Europe project INSTINCT under grant agreement number 101139161, as well as by the Agence Nationale de la Recherche (ANR) through the France 2030 project ANR-PEPR Networks of the Future under grant agreement NF-PERSEUS 22-PEFT-004, and by the CHIST-ERA project PA
}
    }
	\author{Mehdi~Monemi, \textit{Member, ~IEEE}, Mehdi~Rasti, \textit{Senior~Member,~IEEE}, Arthur~S.~de~Sena, \textit{Member~IEEE}, Mohammad~Amir~Fallah, Matti~Latva-aho, \textit{Fellow,~IEEE}, Marco~Di~Renzo, \textit{Fellow,~IEEE},
 \thanks{
  M. Monemi,
  A. Sousa de Sena, and M. Latva-aho are with Centre
for Wireless Communications (CWC), University of Oulu, 90570 Oulu, Finland (emails: mehdi.monemi@oulu.fi, arthur.sena@oulu.fi, matti.latva-aho@oulu.fi).
}
\thanks{M. Rasti is with Centre for Wireless Communications (CWC), and  Water, Energy and Environmental Engineering Research Unit (WE3), University of Oulu, 90570 Oulu, Finland (email: mehdi.rasti@oulu.fi).}
\thanks{
M. A. Fallah is with Department of Engineering, Payame Noor University (PNU), Tehran, Iran (email: mfallah@pnu.ac.ir).
}
\thanks{M. Di Renzo is with Universit\'e Paris-Saclay, CNRS, CentraleSup\'elec, Laboratoire des Signaux et Syst\`emes, 3 Rue Joliot-Curie, 91192 Gif-sur-Yvette, France. (marco.di-renzo@universite-paris-saclay.fr), and with King's College London, Centre for Telecommunications Research -- Department of Engineering, WC2R 2LS London, United Kingdom (marco.direnzo@kcl.ac.uk)}
%\thanks{
%Marco Di Renzo is with Université Paris-Saclay, CNRS, CentraleSupélec, Laboratoire des Signaux et Systèmes, 3 Rue Joliot-Curie, 91192 Gif-surYvette, France. (e-mail: marco.di-renzo@universite-paris-saclay.fr)
%}
}

	\maketitle
% 	\vspace{-15mm}
	\begin{abstract}

Reconfigurable intelligent surfaces (RISs) have been introduced as arrays of nearly passive elements with software-tunable electromagnetic properties to dynamically manipulate the reflection/transmission of radio signals. Research in this area focuses on two applications, namely {\it user-assist} RIS aiming to tune the RIS to enhance the quality-of-service (QoS) of target users, and the {\it malicious} RIS aiming for an
attacker to degrade the QoS at victim receivers through generating {\it intended} destructive interference. While both user-assist and malicious RIS applications have been explored extensively, the impact of RIS deployments on imposing {\it unintended} interference on various wireless user-equipments (UEs) remains underexplored.
This paper investigates the challenges of integrating RISs into multi-carrier, multi-user, and multi-operator wireless networks. We discuss how RIS deployments intended to benefit specific users can negatively impact other users served at various carrier frequencies through different network operators. While not an ideal solution, we discuss how ultra-narrowband metasurfaces can be incorporated into the manufacturing of RISs to mitigate some challenges of RIS deployment in wireless networks. We also present a simulation scenario to illustrate some practical challenges associated with the deployment of RISs in shared public environments.	
%Existing research often focuses on single-carrier scenarios, neglecting the complexities of real-world networks with diverse services, operators, and frequencies.
	\end{abstract}
	%...................................................................................................................................
	% keywords
	
	%\IEEEpeerreviewmaketitle
	
	%...................................................................................................................................
	% Introduction
\thispagestyle{empty}

\section{Introduction}

\begin{textblock*}{10cm}(10.4cm,1cm)  
   Accepted to be published in IEEE Communications Magazine
\end{textblock*}

Reconfigurable intelligent surfaces (RISs) are emerging as a technology to enhance the quality of wireless communication by dynamically manipulating the reflection/transmission of radio signals. Engineered with a large number of nearly passive elements with software-tunable electromagnetic properties, these programmable surfaces manipulate radio waves, offering the potential to impact the signal-to-interference-plus-noise-ratio (SINR) for target users. 
Research in this area has primarily focused on two key applications of RISs. The first application, referred to as {\it user-assist} RIS, aims to improve the quality of communication \cite{9627818} or energy transfer \cite{9743355} between a transmitter and intended receiver(s), particularly when a direct line-of-sight (LoS) path is unavailable and the direct channel gain experiences high path loss due to obstructions. This is accomplished by intelligently directing incident signals at the RIS toward intended users \cite{10077119,9771330}. Conversely, the second application, known as {\it malicious} RIS, involves the deliberate manipulation of radio waves to disrupt targeted users' communication links during data transmission or channel estimation. In such adversarial scenarios, an attacker controls the RIS to degrade the SINR at victim receivers and weaken their quality of service (QoS) \cite{9789438}.

While both beneficial and disruptive applications of RISs have been explored extensively, the impact of RIS deployments on imposing {\it unintended} interference on various wireless user-equipments (UEs) remains underexplored. This is particularly crucial in real-world scenarios where a variety of wireless services operating at different carrier frequencies and potentially served by different and independent operators\footnote{Throughout this paper, the term {\it operator} is used to encompass both licensed network operators (e.g., mobile network operators) and entities responsible for the operation of unlicensed networks (e.g., WiFi network administrators).} might coexist in close proximity to deployed RISs. Existing research often focuses on deploying RISs for a single service provider at a specific carrier frequency, neglecting the potential side effects on other users operating on different frequencies.

Recently, some works have studied the implementation of RISs in multi-operator/provider networks to manage the resource allocation of RIS-assisted UEs \cite{10504275,de2024beyond}. The general idea is to optimize the scattering matrix of the RIS through a multi-objective optimization problem where an aggregate/joint performance measure relating to UEs from several operators is optimized, taking into account the frequency-dependent QoS requirements of both target and non-target UEs. However, these studies face significant limitations. Firstly, they consider a small number of UEs operating within specified service types. Consequently, there is no provision to accommodate a diverse range of service types for various UEs each operating at a dedicated frequency band. Additionally, these methods necessitate synchronized channel estimation for UEs serviced by different operators, and moreover, all channel state information (CSI) between transmitters/base stations (BSs), RISs, and UEs must be universally accessible across the network. These requirements are essential to ensure that the control of the RIS scattering matrix, along with beamforming and resource allocation for all UEs served by various operators, is managed efficiently by a centralized processing unit (CPU) involving all network operators.
Providing access to all BS-RIS and RIS-UEs' channels for a local CPU in an isolated single-operator model is still a challenging issue, let alone for the more complex multi-carrier multi-operator scenarios. Finally, including all non-target UEs in the optimization process for target UE/UEs degrades the QoS of those UEs previously optimized independently.

The aforementioned assumptions and limitations impose practical challenges for RIS deployment in public multi-band wireless environments. For instance, in the vicinity of deployed RISs, there may be UEs utilizing a variety of standardized technologies that exploit licensed or unlicensed spectrum, such as Bluetooth, WiFi, xG cellular services, and others, where the RISs reflect the signals of target and non-target frequencies. Incorporating inter-operator RIS considerations into these technologies to minimize the adverse effect on non-target frequencies is not feasible, as these technologies execute resource allocation at their dedicated frequency bands independently of others based on their own QoS requirements and standardized protocols and restrictions. Considering these limitations, we argue that the implementation of RIS-assisted resource allocation schemes investigated in the literature might present practical challenges in public networks, which are commonly overlooked in research studies.

The contributions and the structure of this work are summarized as follows:
\begin{itemize}
    \item 
    %First, we consider a simplified RIS-assisted multi-user single-carrier network model, disregarding RIS scattering effects on other frequencies. For this scenario, we argue that the RIS aiming to restore the QoS of UEs with blocked connections, potentially adversely affecting other UEs with previously reliable connections. The reasons are discussed in Section II.
    
    %First, we consider a simplified RIS-assisted multi-user {\it single-carrier} network model, ignoring the effects of RIS scattering on other carrier frequencies. We assume that the RIS is intended to restore the QoS of a subset of UEs that have lost reliable connections due to issues such as blockage.     For this scenario, We argue on practical reasons on  how tuning the RIS to restore the QoS of some target UEs can  adversely affect other UEs who were previously enjoying reliable connections before the activation of RIS. The reasons behind this are discussed in Section II.
    First, we consider a simplified RIS-assisted multi-user {\it single-carrier} network model, ignoring RIS scattering effects on other frequencies. We assume the RIS aims to restore the QoS of UEs with blocked connections. However, we argue that tuning the RIS for these target UEs can adversely affect others with previously reliable connections. The reasons are discussed in Section II.
      
    \item
     We investigate the potential practical adverse effects of deploying RISs in more realistic {\it multi-band, multi-operator} networks, where diverse UEs operating on various frequencies %and serviced by different operators
     coexist in proximity to RISs. We also explore the potential of leveraging RISs with ultra-narrowband metasurfaces to mitigate some of these challenges, where their associated benefits and drawbacks are discussed. 
     These are explored in Sections III and V. %The adverse effects of employing RISs in more realistic multi-band multi-operator network environments, where various UEs operating on different carrier frequencies and serviced by different operators coexist near RISs, are investigated in Sections III and IV. We will discuss practical concerns associated with both conve To narrow the practical challenges, employing RISs leveraging ultra-narrowband metasurfaces is suggested as a partial solution whose benefits and drawbacks will be investigated in Section III.

    \item In Section V, we present a simulation scenario for multi-band multi-operator networks demonstrating the dual nature of RISs; while they enhance the QoS for the intended UEs serviced by a given operator at a target frequency, they degrade the QoS for other operators' UEs.% working at other frequencies. 
    
    \item Finally, we conclude the paper by emphasizing that safe and reliable deployment of RISs in public multi-band multi-operator environments potentially requires a careful design scheme and robust regulatory considerations.

\end{itemize}

\begin{figure*}
    \centering
\includegraphics[width=512pt]{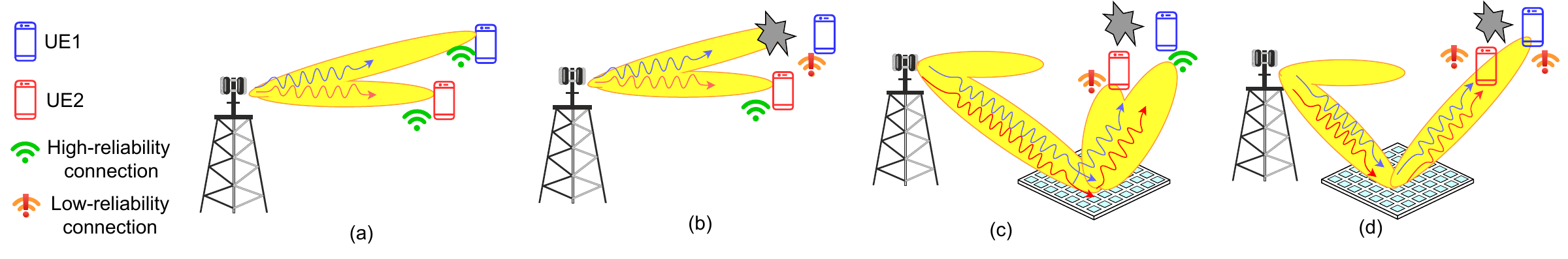}
    \caption{Single-operator two-user MIMO communication performed either through precoding-only or aided by an RIS. (a): Both UEs have LoS-dominant links and reliable connections without involving an RIS. (b): LoS path for UE1 is blocked.  (c): LoS path for UE1 is blocked. The RIS is tuned to meet the minimum QoS requirement of UE1, which might unintentionally degrade the QoS of UE2 due to several factors. % are respectively provided with high-reliability and low-reliability connections. 
    (d): Los path for UE1 is blocked and UE2 is placed between the RIS and UE1. Both UEs experience low-reliability connections due to a common dominant path and highly correlated channels. Here, the single-carrier MIMO precoding becomes ineffective.}
    \label{fig:single_carrier}
\end{figure*}
\section{RIS-Induced QoS Challenges in Single-Carrier Multi-User Networks}
\label{sec:single_carrier}
Single-carrier multi-user communication is the common scenario studied in most RIS-aided studies in the literature, wherein the impact of the RIS on frequencies other than that of interest is ignored. Even this simplified model still pertains to some challenges in practice. For instance, consider a network where some UEs cannot establish reliable connections due to blockage. Let us define the infeasibility measure as the number of such UEs. We will argue that activating an RIS to decrease the infeasibility measure is not always possible. %, i.e., while it may improve the QoS for blocked UEs, it might inadvertently degrade the QoS of other UEs with previously reliable connections. 
While some studies on RIS-assisted underlay cognitive radio (CR) networks have considered mitigating unintended RIS-induced interference on primary users \cite{9776539}, \cite{9146170}, %through techniques like interference nulling \cite{9776539} or temperature-limit-based \cite{9146170} approaches,
we delve deeper into the underlying reasons for reliability issues and potential performance degradation in both CR and non-CR RIS-assisted underlay networks, primarily from a channel modeling perspective. %Consider a downlink communication scenario where a BS transmits data streams to $M$ UEs. 
%Existing research on RIS-aided resource allocation typically optimizes an objective function (e.g., sum rate) by adjusting the precoder at the BS and the scattering matrix (corresponding to the phase shifts) of the RIS. 
Considering a RIS-aided network optimization problem, guaranteeing a minimum QoS level for each UE becomes crucial when reliable links are essential.  This translates into a set of feasibility constraints for the UEs that must be incorporated into the main optimization problem, where the satisfaction of all such constraints might not always be feasible in all network models and environments.

Fig.~\ref{fig:single_carrier}-a depicts a scenario showing a BS serving two UEs where LoS paths exist for both, and the channel vectors of the two UEs are not highly dependent.  In this case, the interference of each UE on the other can be effectively canceled and an acceptable QoS level can be obtained by designing suitable precoders without requiring the deployment of RIS.
Now assume that the LoS connection of UE1 is blocked, as seen in Fig.~\ref{fig:single_carrier}-b, leading to the QoS degradation of that UE. 
An RIS previously installed at some place in the network can be activated to restore the desired QoS of UE1 as shown in Fig.~\ref{fig:single_carrier}-c. However, optimizing the RIS scattering matrix and the BS precoder to guarantee UE1's reliability might come at the expense of UE2's loss of reliability. This might occur due to several reasons:
(a) {\it Reduced Direct Power:}  To prioritize UE1's QoS, the BS transmits less power directly towards UE2, (b) {\it Incoherence:}  The direct radiated power toward UE2 and the reflected power from the RIS might experience some degree of incoherence, further degrading the signal quality for UE2, and finally
(c) {\it Interference:}  The reflected power intended for UE1 can also reach UE2;  Interference mitigation techniques such as zero-forcing can sometimes be challenging in practice due to the geometrical positions of the UEs relative to the RIS. The last issue is more elaborated on in the following.
Consider a scenario depicted in  Fig.~\ref{fig:single_carrier}-d where both UEs are placed at relatively same direction angle with respect to the RIS. It is seen here that both UEs might be negatively affected if the RIS is utilized.
The uncorrelated BS-UEs' channel vectors (and consequently the full-rank BS-UEs channel matrix) employed for efficient multi-user MIMO communication in Fig.~\ref{fig:single_carrier}-a are undermined by the highly correlated near singular RIS-to-UEs' channels matrix in Fig.~\ref{fig:single_carrier}-d due to UE2 being located in the LoS dominant path between the RIS and UE1. This situation makes it challenging to design efficient MIMO precoders to eliminate interference between the UEs, potentially leading to QoS reliability issues for both UE1 and UE2.
These factors highlight the trade-offs inherent in RIS deployment for single-carrier multi-user networks. %While it may improve the communication quality for a target user, it can also negatively impact the QoS of other users.
To mitigate this, further techniques might be required to be incorporated, including RIS-aware rescheduling and carrier assignment. For example, a solution to resolve the precoder design issues in scenarios like Fig.~\ref{fig:single_carrier}-c and Fig.~\ref{fig:single_carrier}-d is to assign independent precoders by considering different orthogonal carriers allocated to UE1 and UE2. However, the multi-carrier scenario introduces other challenges, which are investigated in the following section. 

\section{
Reasons Behind Reliability Risks in Multi-Carrier Multi-Operator Networks}

\label{sec:Multicarrier_Reasons}
The reliable deployment of RISs becomes more challenging in practice where a variety of UEs operate in different frequency bands and are served by various operators. Hereafter, we use the term {\it target UE/UEs} to the user(s) for which the RIS's scattering matrix is to be tuned. All other UEs existing in the vicinity of the RIS are referred to as {\it non-target UEs}. The corresponding carrier frequencies are respectively referred to as {\it target} and {\it non-target} carrier frequencies. The reliability hazards of an RIS in real-world multi-carrier heterogeneous networks stem from two important properties investigated in the following subsections.

\subsection{Frequency Response of RISs}

From a system-level perspective, RISs operate similarly to active relays forwarding the transmission beam toward target UEs.
However, relays utilize various active and passive filtering levels to ensure that only the signal with the intended carrier frequency is processed and forwarded. Here, the effective frequency response is the cascaded effect of several filters including wideband frequency responses of the relay's receive and transmit antennas, mid-bandwidth bandpass analog filters such as the ones for eliminating out-of-band noise, and finally, sharp low-bandwidth digital filters for fine-tuning and passing only the intended low-bandwidth signal at the target carrier frequency. 
In contrast, the frequency domain filtering of RISs is limited to the inherent wideband frequency response of the reflecting surfaces, which allows a broad range of carrier frequencies associated with other UEs and networks to pass through. The range of frequencies that an RIS can impact signal propagation in either reflective or transmissive mode is defined as the bandwidth of influence (BoI) \cite{alexandropoulos2023ris}. Table I compares the fractional BoI (the ratio of BoI over the central frequency) for 4 different types of typical RIS unit cells validated through experimental results in the European Union Horizon 2020 RISE-6G project \cite{alexandropoulos2023ris}.
To compare the fractional BoI data in Table I with the bandwidth required for a typical cellular UE connection, consider a 5G network operating at the FR1 1.8 GHz frequency, and assume that a resource block of 12 consecutive subcarriers each having a 30 kHz subcarrier spacing is completely allocated to some UE.  Here, the total allocated bandwidth is $360$ KHz and the corresponding fractional bandwidth is $0.02$\%. Comparing this with the values expressed in Table I reveals how significantly the fractional BoI of RISs exceeds what is required by a typical 5G target UE. 
 While narrowing the bandwidth of reflective/transmissive RISs can be somewhat achievable through specialized RIS's elements design, this approach can compromise the surface's reflectivity/transmissivity and, in any case, cannot replicate the sharp digital filtering and mid-bandwidth analog filters employed in active relays.
 
\begin{table}[!t]
 \caption{Comparison of the BoI and fractional BoI of 4 typical RISs studied through experimental results for RISE-6G project.}
     \centering
    \begin{tabular}{|l|l|l|} \hline 
            {\bf RIS Type}&  {\bf BoI (GHz)}& {\bf BoI/$\boldsymbol{f_0}$}\\ \hline 
           {1-bit transmissive}&  [23.9\ 30.6]& 24.5\%\\ \hline 
           {varactor-based reflective}&  [5.1\ 6.4]& 22.4\%\\ \hline 
            RF-switch-based transmissive &  [5.17 5.44]& 5.1\%\\ \hline
  PIN-diode-based transmissive& [23.9\ 30.6]&23.7\%\\\hline
    \end{tabular}
    \label{tab:my_label}
\end{table}

In recent years, the fabrication of ultra-narrowband (UN) metasurface antennas has been investigated, where the antenna bandwidth can be reduced to less than 1 percent of the carrier frequency through employing special kinds of metamaterials (e.g., \cite{shangguan2022design}). While such metasurfaces have been explored and fabricated as transmit and receive antennas, they can potentially be incorporated into RISs as well. However, this integration poses several challenges. Firstly, the manufacturing complexity and cost of UN metasurfaces are much higher than those of ordinary metasurfaces, especially when designed in the form of arrays with many meta-atoms. This complexity contradicts one of the main goals of RISs, which is to be cost-effective and easily mountable. Secondly, in practical wireless communications, single/multiple carrier frequencies from a pool of carriers in the available spectrum are dynamically allocated to target UEs. As a result, the designed RIS must have a bandwidth encompassing all carriers in the available spectrum, which inevitably leads to the reflection of all signals with carrier frequencies allocated to non-target UEs being served in the available spectrum.

\subsection{Beam Squint: A Destructive Issue Associated with RIS-Aided Networks}
\label{sec:squint}
  Beam squint is a known phenomenon in beamforming using phased-array antennas. This generally refers to the deviation of the main beam direction when the radiated signal frequency spans over a large spectrum as is the case in wideband signal applications. The investigation of beam squint in non-RIS-aided and RIS-aided beamforming applications has been addressed in many works (e.g., \cite{10130575,9771341}), however, these works consider the beam squint for the case where a wideband target signal is radiated from an intended transmitter. 
  In practice, however, irrespective of the bandwidth of the target signal, due to the wide BoI of RISs and the non-linear frequency-dependent scattering matrix (discussed later in this section), the beam squint is unavoidable for signals of non-target UEs operating at frequencies within the BoI of the RIS.  This means that the RIS scatters unintended signals with different frequencies in various directions,
  potentially resulting in uncontrolled interfering effects for various non-target UEs operating around the RIS. 
  
\begin{figure}
    \centering
\includegraphics[width=254pt]{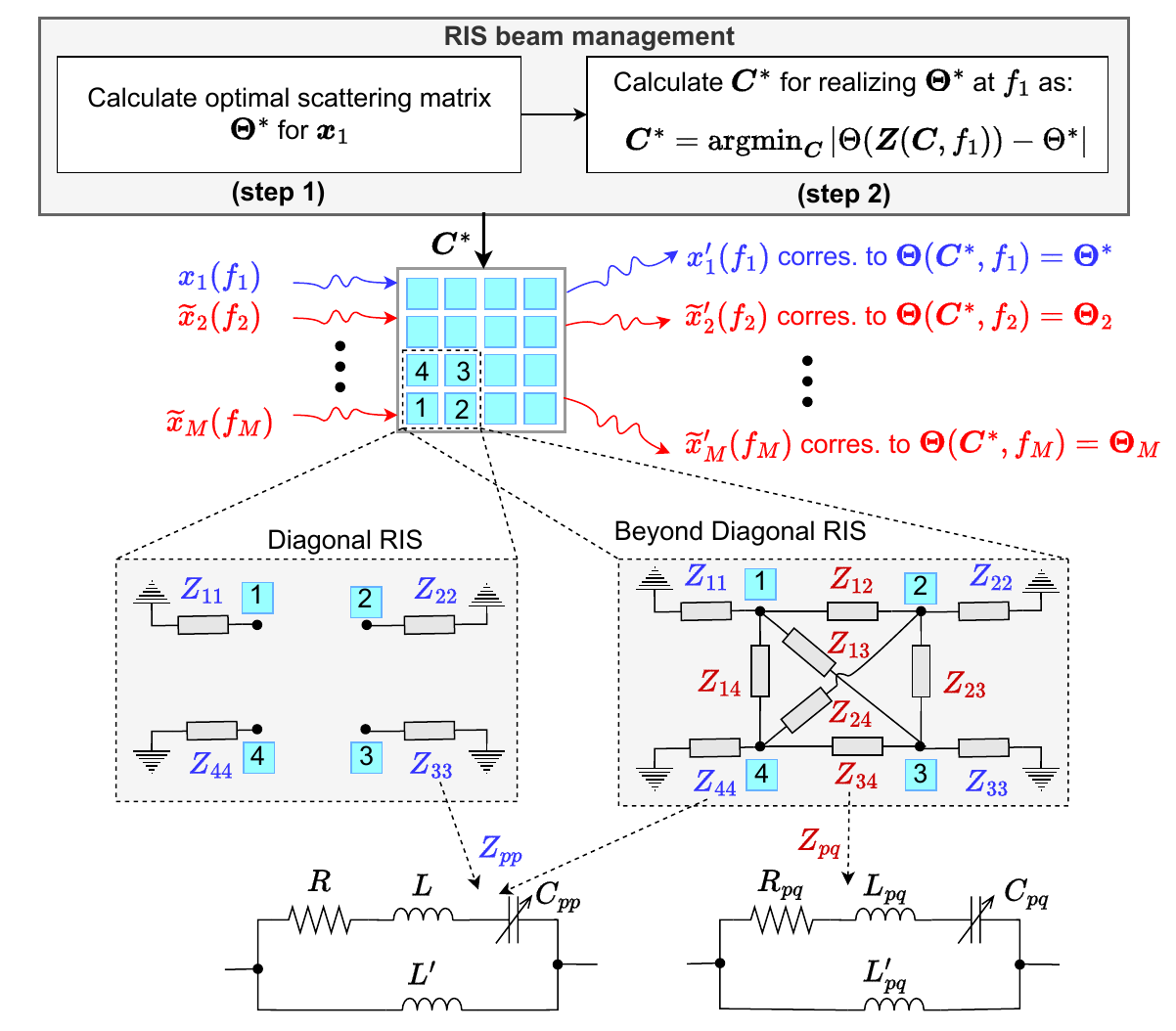}
    \caption{ The mechanism for tuning the scattering matrix for a target signal $x_1$ at frequency $f_1$ resulting in the scattering of non-target signals in different directions. A section of the circuit model of reflective elements for diagonal and beyond diagonal RISs is also illustrated.}
    %\caption{System model and scattering matrix tunning for RISs. (a): A section of the circuit model of reflective elements for diagonal and beyond diagonal RISs. (b): The mechanism for tuning the scattering matrix for target signal $x_1$ at frequency $f_1$ resulting in the scattering of non-target signals in different directions.}
    \label{fig:scattering_model}
\end{figure}

  The element-wise frequency-dependent radiation of metasurfaces has been explored in some works (e.g., \cite{9321220}). To extend this from a system-level phased-array perspective, we investigate an RIS composed of $N$ reflective meta-atoms. Let's consider a varactor-based RIS as one of the most popular types employed in practice, where the reflective behavior can be characterized by the circuit model depicted in Fig.~\ref{fig:scattering_model}  \cite{9759366,de2024beyond}. %This model can be modified to represent other types of RISs.
  The reflection pattern of the RIS is determined by tuning the scattering matrix $\boldsymbol{\Theta} \in \mathbb{C}^{N\times N}$. Considering the circuit model of RIS elements depicted in the figure for each of the diagonal and beyond diagonal models, each realized impedance matrix $\boldsymbol{Z} \in \mathbb{C}^{N\times N}$ results in some scattering matrix $\boldsymbol{\Theta}$.
In either case, as depicted in the RIS beam management block in Fig.~\ref{fig:scattering_model}, practical RIS beam tuning for a set of target UEs operating at a target carrier frequency $f_1$ involves the following steps:

    {\bf Step 1:}
    Considering a given system model and problem formulation, 
    obtain the optimal scattering matrix $\boldsymbol{\Theta}_1=\boldsymbol{\Theta}^*$  to reflect the desired signal $x_1$ optimally toward target UEs operating at frequency $f_1$.
    
    {\bf Step 2:} Based on the circuit model of the RIS shown in Fig. \ref{fig:scattering_model}, and noting that the impedance matrix $\boldsymbol{Z}$ and consequently the scattering matrix $\TB$ is a function of the frequency $f$ and capacitance matrix
$\boldsymbol{C}$, calculate the capacitance matrix $\boldsymbol{C}^*$ such that $\TB^*$ is realized with minimum error. This corresponds to $ \boldsymbol{C}^*=\argmin_{\boldsymbol{C}} \left| \TB(\boldsymbol{Z}(\boldsymbol{C},f_1)) - \TB^* \right|$. This can be achieved in a way described in \cite{9759366} and \cite{de2024beyond} for diagonal and beyond-diagonal RIS models, respectively.

\begin{figure}
    \centering
\includegraphics[width=234pt]{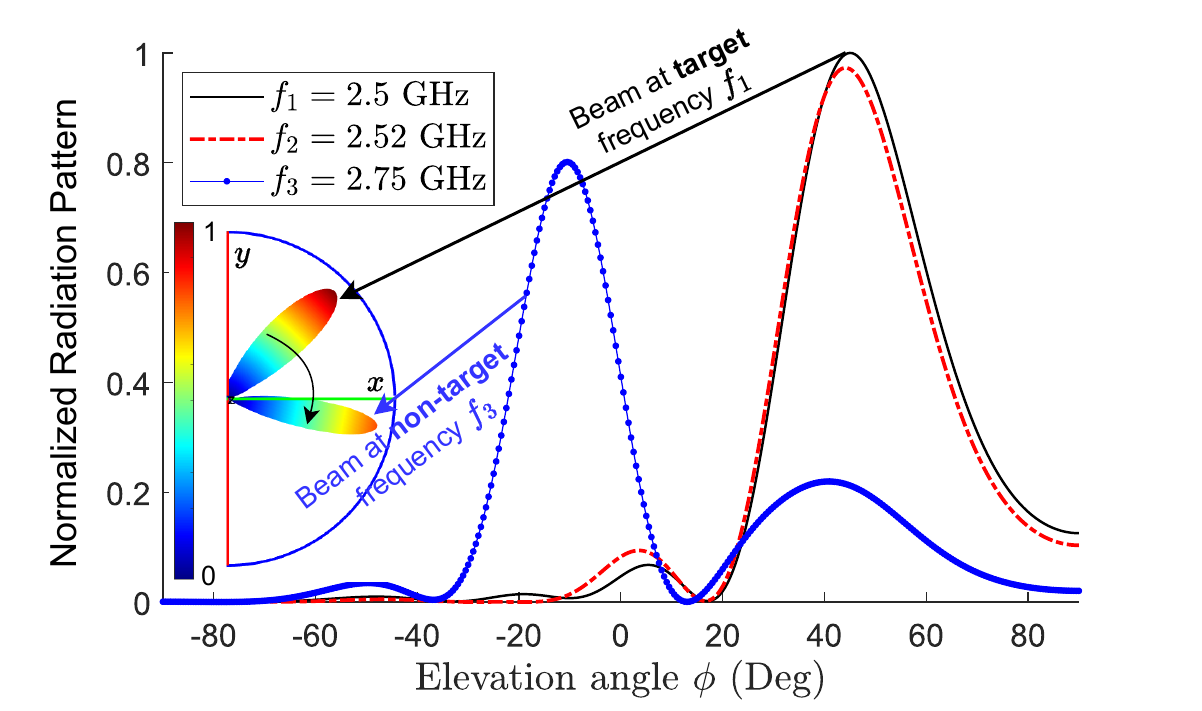}
    \caption{Demonstration of how the RIS reflection pattern changes for signals with different frequencies. A BS is located at $(50,-50, 18)$m transmitting signals at three frequencies $f_1$, $f_2$ and $f_3$. The target UE operating at $f_1$ is located at $(3,3,0)$m. A $20\times 20$ RIS is placed on the $xz$ plane, centered at the origin with an inter-element spacing of half a wavelength.}
    \label{fig:directivity}
\end{figure}

Having determined the capacitance matrix $\boldsymbol{C}^*$ optimized for the target frequency according to {\bf Step~2}, the scattering matrix  $\TB(\boldsymbol{Z}(\boldsymbol{C}^*,f)) \equiv \TB(f)$ is found to be directly a function of the carrier frequency $f$. Noting that a wide range of non-target signals with carrier frequencies $f_m\neq f_1$, $m \in \{2,3,...,M\}$ within the BoI might exist in the vicinity of the RIS, the non-target incident signals $\widetilde{x}_m(f_m),\forall m$ experience different squints and are reflected in different directions, each based on its scattering matrix  $\boldsymbol{\Theta}_m=\boldsymbol{\Theta}({\boldsymbol{C}^*}, f_m)$ and the corresponding angle of arrival (AoA). These reflections can potentially have various detrimental effects on non-target UEs, as will be discussed in the next section.
To visualize this, consider a $20\times 20$ diagonal RIS in a scenario described in Fig. \ref{fig:directivity}. The scattering matrix is configured for the target carrier frequency $f_1 = 2.5$ GHz. The impedance matrix parameters are set according to \cite{9759366}. Assume signals for two other non-target UEs also impinge on the RIS surface at the same AoA, but with carrier frequencies $f_2 = 2.52$ GHz and $f_3 = 2.75$ GHz. Notably, a small frequency deviation from $f_1$ to $f_2$ results in a reflection pattern for the non-target signal $x_2$ that closely resembles that of the target signal $x_1$. In contrast, changing the frequency from $2.5$ GHz to $2.75$ GHz leads to a significant deviation in the reflection direction from $45^{\circ}$ to $-14^{\circ}$, corresponding to the target and non-target reflected beams respectively. Note that unlike the conventional study of beam squint for wideband target signals, here the existence of beam squint for any traveling signal at carrier frequency $f_m \neq f_1$ is not contingent on whether the target signal $x_1$ is narrowband or wideband, rather, the BoI which is commonly wideband is the determining factor.% However, the beam squint is not the case if the RIS comprises ultra-narrowband metasurface elements.
 \section{
 %RIS Deployment in Heterogeneous Multi-Carrier Multi-Operator Networks: Scenarios Resulting in Reliability Risks
 Scenarios Resulting in Reliability Risks in Multi-Carrier Multi-Operator Networks
 }
 \label{sec:Multicarrier_Scenarios}
 Building upon the inherent issues of nearly passive RIS technology discussed in the previous section, in what follows we examine specific scenarios and forms where these issues can lead to unintended QoS consequences for non-target UEs.

\subsection{Destructive Self-Interference}
%Recall that the RIS reflects signals from non-target UEs. These might serve as destructive interference to such UEs. Fig. \ref{fig:multi_carrier} depicts several scenarios where multiple UEs exist in a network, and the RIS scattering matrix is tuned for the target signal $x_1$ at frequency $f_1$. First note that if the non-target frequency $f_2$ is close to $f_1$ the reflected signal experiences a negligible beam squint (Fig. \ref{fig:multi_carrier}-a) and otherwise the main lobe of the reflected signal $x'_2$ is in a different direction from that of the target signal. In either case, if the non-target UE is not located in the main lobe of the corresponding reflected signal (Fig. \ref{fig:multi_carrier}-a1 and Fig. \ref{fig:multi_carrier}-b1), it will be protected, otherwise (Fig. \ref{fig:multi_carrier}-a2 and Fig. \ref{fig:multi_carrier}-b2), it is potential to QoS degradation due to potentially incoherent multi-path (direct and reflected) propagation of the signal for UE2 as the non-target user. Although the RIS can be tuned in a way that the signal from the direct path (if not blocked) and the reflected path arrive constructively at the UE, this however holds only for the target UEs. Fig. \ref{fig:multi_carrier}-c2 illustrates a similar problem when two service providers exist in the network.
Recall that RISs also reflect the signals transmitted for non-target UEs. These reflections can impose destructive interference on corresponding 
UEs. To showcase some examples, Fig. \ref{fig:multi_carrier}-a and \ref{fig:multi_carrier}-b illustrate scenarios in a network where the RIS scattering matrix is to be tuned for the target user $\mathrm{U}_1^{(1)}$ operating at frequency $f_1$, which in turn affects the non-target user $\mathrm{U}_2^{(1)}$ operating at frequency $f_2$. 
Depending on the channel model assumption, various harmful self-interference effects for the non-target UE can occur.  Generally, the reflected rays from each RIS element toward the non-target UE combine with the signal received from the direct path, forming a multi-path fading channel. If the LoS channel model is considered, the beam pattern shaped by the RIS and potentially directed toward the non-target UE might increase the interference.
If the non-target frequency $f_2$ is close to $f_1$, the reflected signal experiences negligible beam squint (Fig. \ref{fig:multi_carrier}-a). Otherwise, the main lobe of the reflected signal $x'_2$ points in a different direction from that of the target reflected signal (Fig. \ref{fig:multi_carrier}-b). In either case, when the non-target UE lies within the main lobe of the reflected signal as shown in Fig. \ref{fig:multi_carrier}-a and Fig. \ref{fig:multi_carrier}-b, QoS degradation is likely to happen due to the potential incoherent multi-path arrival of the signal rays. While the RIS can be tuned to ensure the constructive arrival of signals from the direct path and reflected path, achieving the desired QoS for target UEs may not result in the desired QoS for the non-target UEs. %Fig. \ref{fig:multi_carrier}-c2 illustrates a similar issue when two service providers coexist in the network and the RIS enforces self-interference for UE2. For the case of multi-operator MIMO communications, as depicted in Fig. \ref{fig:multi_carrier}-c, tuning the RIS for target UEs at the target frequency results in the channel alteration for all non-target MIMO UEs, leading to unintended beam squint and self-interference for all such UEs. This further results in precoders  

%\subsection{Post-Precoding Error}
%This is a direct result of the self-interference in multi-operator MIMO communications, where each BS is supposed to serve multiple MIMO links at each dedicated carrier frequency, as seen in Fig. \ref{fig:multi_carrier}-c. Once the RIS is tuned for a set of UEs operating at a given frequency, the precoders 

\subsection{CSI Perturbation and Inaccuracy of Precoding }
The scattering matrix of the RIS is continuously and frequently updated based on the CSI update schedule, considering changes in both BS-to-RIS and RIS-to-target-UE channels.  This update process, however, does not account for potential negative impacts on non-target UEs. The CSI estimation schedules of non-target UEs, which operate under different network protocols and are managed by separate network coordinators, are not synchronized with those of the target UEs. As a result, the estimated channels of non-target UEs become outdated as soon as the RIS scattering matrix is reconfigured for the target UEs. This desynchronization reduces the channel coherence time, making it unpredictable and degrading the signal quality and connection reliability experienced by non-target UEs. For multi-operator MIMO communications, where each BS is supposed to serve multiple MIMO links at each dedicated carrier frequency, as depicted in Fig. \ref{fig:multi_carrier}-c, this together with the destructive self-interference for non-target UEs results in increasing decoding error probability due to inaccurate and outdated computed precoders. %Once the RIS is tuned for a set of UEs operating at a given frequency, the precoders 

\subsection{Destructive Cross-Interference}
\begin{figure}
    \centering
\includegraphics[width=254pt]{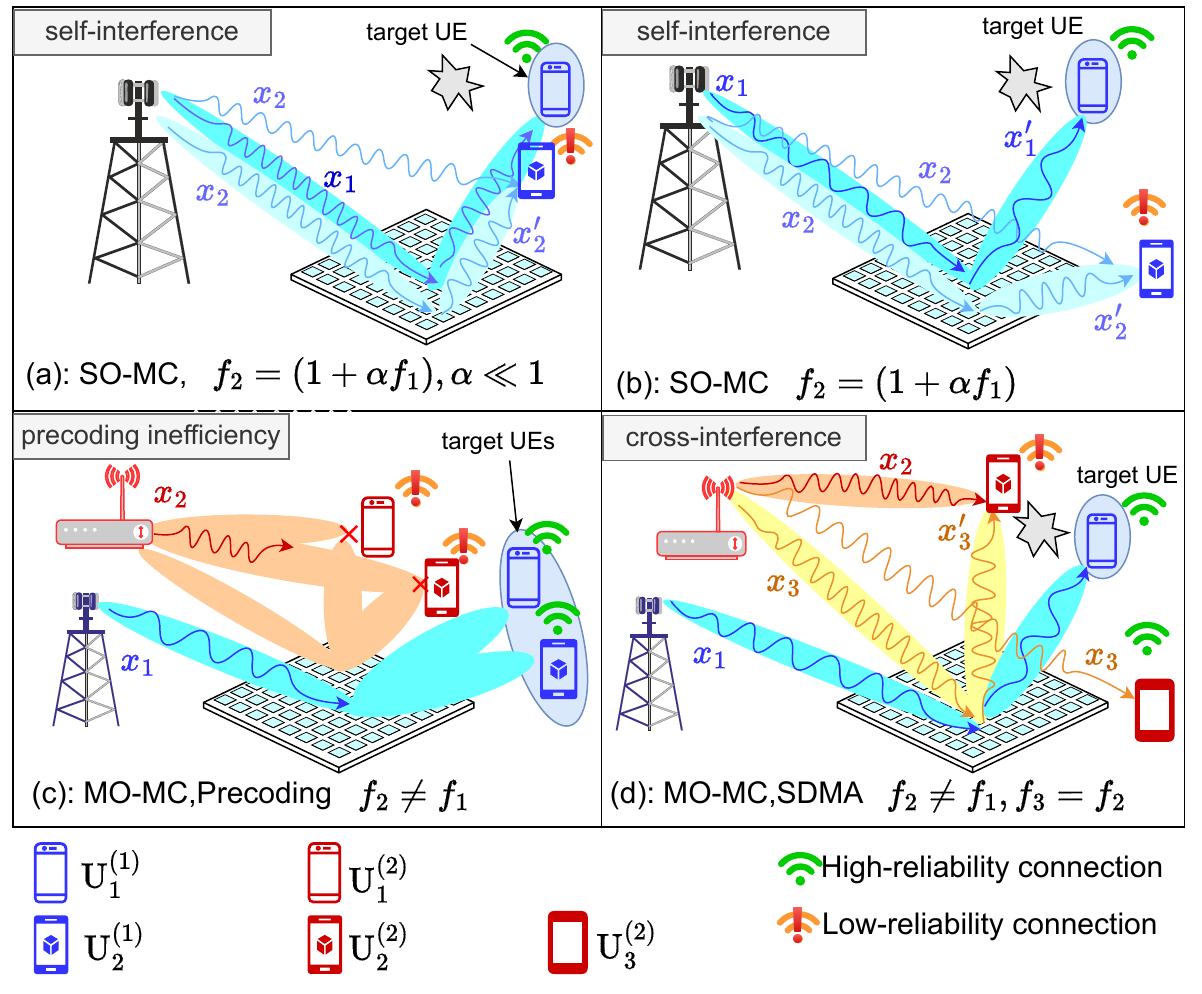}
    \caption{Illustration of various self-interference and cross-interference scenarios for non-target UEs in a single-operator multi-carrier (SO-MC), and multi-operator multi-carrier (MO-MC) network, where $\mathrm{U}_i^{(m)}$ denotes the $i$-th user of AP $m\in\{1,2\}$. }%$\mathrm{U}_1^{(1)}$ and $\mathrm{U}_2^{(1)}$ are respectively the first and second UEs communicating with AP1, and $\mathrm{U}_1^{(2)}$, and $\mathrm{U}_2^{(2)}$ are the first and second non-target UEs communicating with AP2.}
    \label{fig:multi_carrier}
\end{figure}

This occurs when the RIS disrupts the spatial domain orthogonality of channel access for the UEs operating at the same carrier frequency. Consider a communication scenario shown in Fig. \ref{fig:multi_carrier}-d where $\mathrm{U}_1^{(1)}$ is the target user of the RIS and the non-target UEs $\mathrm{U}_2^{(2)}$ and $\mathrm{U}_3^{(2)}$ are spatially separated and served by another access point (AP) 2 on the same carrier frequency using spatial division multiple access (SDMA). Although $\mathrm{U}_2^{(2)}$ and $\mathrm{U}_3^{(2)}$ are considered to be sufficiently separated in the angular domain with respect to the corresponding transmitter to ensure effective SDMA performance, the deployment of the RIS may cause the signal $x_2$ intended for $\mathrm{U}_2^{(2)}$ to be reflected toward  $\mathrm{U}_3^{(2)}$. As a result, cochannel cross-interference for  $\mathrm{U}_3^{(2)}$ becomes a more challenging scenario compared to the previously discussed self-interference scenario. Unlike self-interference, which may occasionally be non-destructive, the cross-interference caused by RIS reflection is consistently destructive.
\section{Numerical Results}
\begin{figure}
    \centering
\includegraphics[width=244pt]{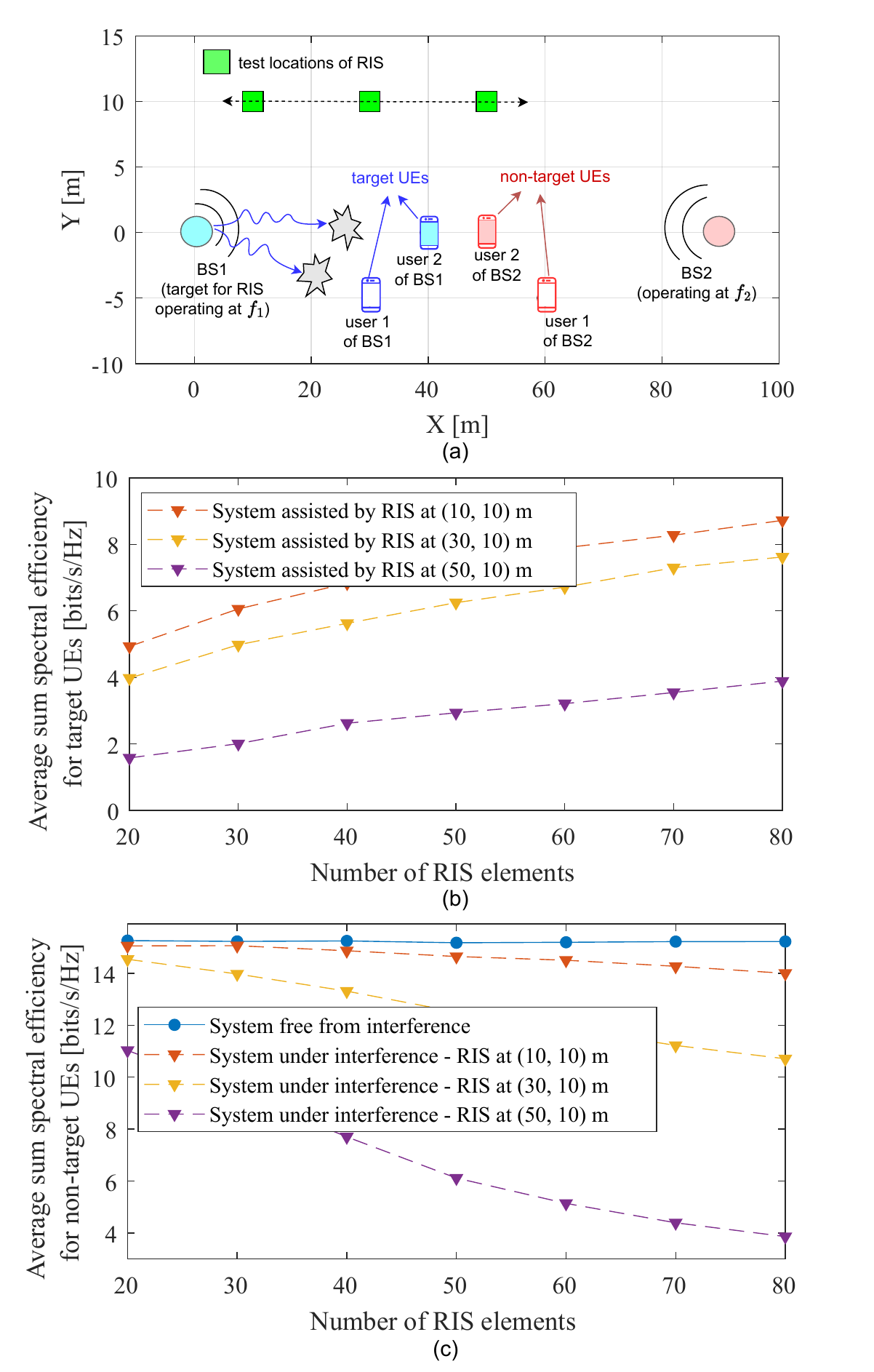}
    \caption{The upper diagram shows the simulation setup scenario considering two BSs from two different operators, where the target and non-target UEs are serviced by operators 1 and 2 at carrier frequencies $f_1=2.5$ GHz and $f_2=2.75$ GHz respectively. The middle plot illustrates the performance measure of RIS target UEs communicating with BS1. The lower plot shows the RIS-induced performance degradation of non-target UEs communicating with BS2.% UEs when the RIS is incorporated for optimizing the performance of target UEs.
    }
    \label{fig:performance_versus_ris_elements}
\end{figure}
In this section, we provide a simulation setup to evaluate the detrimental consequence of RIS deployment in multi-operator networks and show how deploying RISs to benefit the UEs served by a given operator can harm other UEs located in the vicinity of the RIS served by other operators on carrier frequencies within the BoI of the RIS. We considered a MIMO network with two operators each having one BS with $M=20$ antennas. We consider that BS1 and BS2 are served by operator 1 and operator 2 at frequencies $f_1=2.5$ GHz and $f_2=2.75$ GHz, respectively, and an RIS is used to assist the target UEs of BS1 whose LoS links are blocked.
A two-dimensional network scenario is investigated according to Fig. \ref{fig:performance_versus_ris_elements}-a where an $N$-element diagonal RIS is considered to be located on 3 test positions. The RIS scattering matrix is obtained by solving the optimization problem corresponding to the maximum weighted sum power of target UEs, where a zero-forcing BS precoder vector is employed for the UEs of each operator  \cite{de2024beyond}. The RIS impedance matrix is obtained from the diagonal model in Fig. \ref{fig:scattering_model}, where $R=1\ \Omega$, $L=2.5\ \mathrm{nH}$, and $L'=0.7\ \mathrm{nH}$ have been considered. %The channel model, RIS model parameter values, and the solution scheme for finding the scattering matrix are adopted from \cite{de2024beyond}.
We employ a Rician fading channel model with Rician factor $K=1$ and an additive white Gaussian noise with variance $-30$ dBm. Here, we assume that the precoders of the non-target UEs are tuned independently without the involvement of the RIS, as the RIS is owned and controlled by the first operator to benefit its users. The tuning of the scattering matrix for the target UEs operating at $f_1$ results in a change in the channel matrix of the non-target UEs operating at $f_2$, in a mechanism expressed in Section \ref{sec:squint}, which leads to the performance degradation for the non-target UEs due to the increase in decoding error probability. 

The double-edged impact of RIS is evidenced by comparing Figs. \ref{fig:performance_versus_ris_elements}-b and \ref{fig:performance_versus_ris_elements}-c, where, for all scenarios, increasing the number of RIS elements enhances the performance of the target UEs (Fig. \ref{fig:performance_versus_ris_elements}-b), while simultaneously exacerbating the negative impacts on the non-target UEs (Fig. \ref{fig:performance_versus_ris_elements}-c).
It is seen from the figure that the RIS-induced performance degradation is relatively high, especially when the number of RIS elements gets higher and the RIS gets closer to the non-target UEs. For example, considering a 40-element RIS located at (50,10)~m, the average sum spectral efficiency of the non-target UEs is decreased by a factor of 49\% from 15.2 b/s/Hz to 7.7 b/s/Hz corresponding to RIS-interference-free (no RIS) and RIS-induced interference scenarios, respectively. 

\begin{figure}
    \centering
\includegraphics[width=254pt]{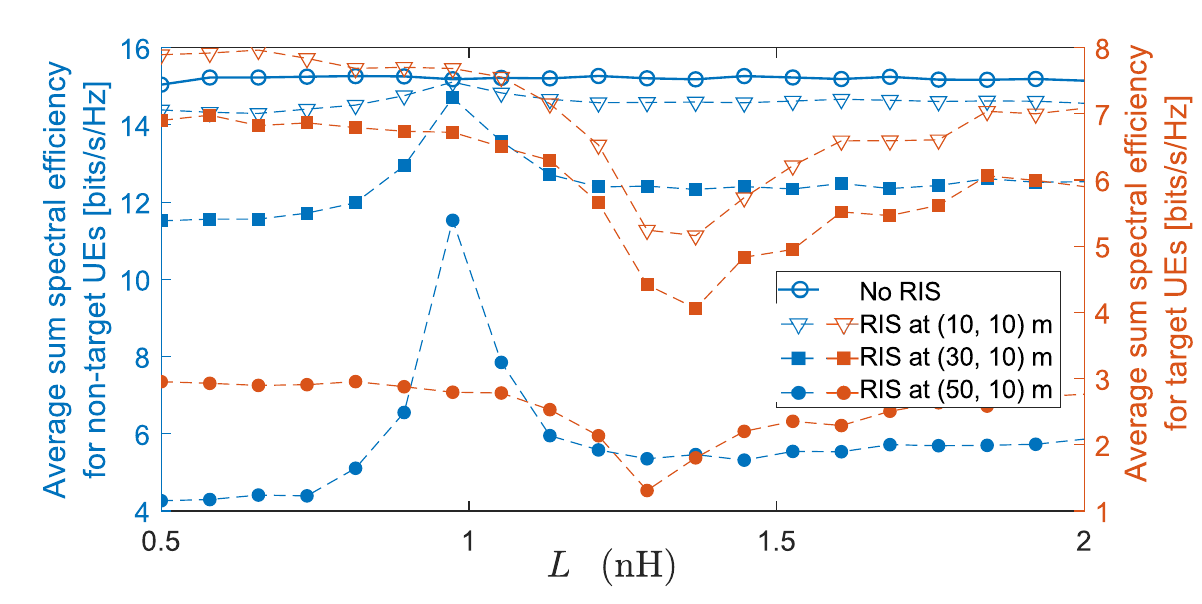}
    \caption{Performance evaluation of target and non-target UEs versus $L$, a system model parameter influencing BoI and beam squint. A 50-element RIS is considered. 
    }
    \label{fig:performance_versus_L}
\end{figure}

As illustrated in the circuit model in Fig. \ref{fig:scattering_model}, each RIS element can be modeled as a bandpass filter, where the variation in the circuit parameter values affects the BoI of the filter and also influences the beam squint for non-target UEs, as elaborated in Section III. 
To investigate the impact of RIS model parameter variations on system performance, we selected inductance $L$ as one of the model parameters and analyzed the performance of both target and non-target UEs for different values of $L$, as shown in Fig. \ref{fig:performance_versus_L}. 
It is seen that the performance variation for both user groups is not monotonic. Furthermore, while better performance for target UEs corresponds to worse performance for non-target UEs for any given model parameter value $L$, the worst and best performance measures for target and non-target UEs, respectively, do not necessarily coincide with the same values of $L$.

%The results indicate that the performance degradation for non-target UEs is consistently lower in the RIS-assisted scenario across all values of L, aligning with the observations in Fig. \ref{fig:performance_versus_ris_elements}. However, 

\section{Conclusion}
In this paper, we explored how RIS deployments intended to benefit specific users can negatively impact other users serviced by different network operators. This underscores the fact that the RIS design and its deployment in public environments require careful consideration. Given the potential impact on users across different networks, a robust regulatory framework is essential to mitigate these adverse effects and establish clear performance metrics, interference limits, and certification requirements for RIS deployments in multi-band multi-operator public network environments. %By establishing clear guidelines for RIS deployment, policymakers can ensure equitable access to wireless resources and ensure that the reliability of the communication for RIS non-target users is not negatively affected by RISs.
%\section{Conclusions}
%In this paper, we uncovered how the deployment of RISs for optimizing the performance of target UEs might result in unintended QoS degradation of other UEs operating in the vicinity of the RIS. This reveals the fact that 
% 	\footnotesize

\bibliographystyle{IEEEtran}
\bibliography{Mybib}

\begin{thebibliography}{10}
\providecommand{\url}[1]{#1}
\csname url@rmstyle\endcsname
\providecommand{\newblock}{\relax}
\providecommand{\bibinfo}[2]{#2}
\providecommand\BIBentrySTDinterwordspacing{\spaceskip=0pt\relax}
\providecommand\BIBentryALTinterwordstretchfactor{4}
\providecommand\BIBentryALTinterwordspacing{\spaceskip=\fontdimen2\font plus
\BIBentryALTinterwordstretchfactor\fontdimen3\font minus \fontdimen4\font\relax}
\providecommand\BIBforeignlanguage[2]{{%
\expandafter\ifx\csname l@#1\endcsname\relax
\typeout{** WARNING: IEEEtran.bst: No hyphenation pattern has been}%
\typeout{** loaded for the language `#1'. Using the pattern for}%
\typeout{** the default language instead.}%
\else
\language=\csname l@#1\endcsname
\fi
#2}}

\bibitem{9627818}
E.~C. Strinati, G.~C. Alexandropoulos, H.~Wymeersch, B.~Denis, V.~Sciancalepore, R.~D'Errico, A.~Clemente, D.-T. Phan-Huy, E.~De~Carvalho, and P.~Popovski, ``{Reconfigurable, Intelligent, and Sustainable Wireless Environments for 6G Smart Connectivity},'' \emph{IEEE Communications Magazine}, vol.~59, no.~10, pp. 99--105, 2021.

\bibitem{9743355}
E.~Shi, J.~Zhang, S.~Chen, J.~Zheng, Y.~Zhang, D.~W. Kwan~Ng, and B.~Ai, ``{Wireless Energy Transfer in RIS-Aided Cell-Free Massive MIMO Systems: Opportunities and Challenges},'' \emph{IEEE Communications Magazine}, vol.~60, no.~3, pp. 26--32, 2022.

\bibitem{10077119}
R.~Liu, M.~Li, H.~Luo, Q.~Liu, and A.~L. Swindlehurst, ``{Integrated Sensing and Communication with Reconfigurable Intelligent Surfaces: Opportunities, Applications, and Future Directions},'' \emph{IEEE Wireless Communications}, vol.~30, no.~1, pp. 50--57, 2023.

\bibitem{9771330}
P.~Wang, J.~Fang, W.~Zhang, Z.~Chen, H.~Li, and W.~Zhang, ``{Beam Training and Alignment for RIS-Assisted Millimeter-Wave Systems: State of the Art and Beyond},'' \emph{IEEE Wireless Communications}, vol.~29, no.~6, pp. 64--71, 2022.

\bibitem{9789438}
Y.~Wang, H.~Lu, D.~Zhao, Y.~Deng, and A.~Nallanathan, ``{Wireless Communication in the Presence of Illegal Reconfigurable Intelligent Surface: Signal Leakage and Interference Attack},'' \emph{IEEE Wireless Communications}, vol.~29, no.~3, pp. 131--138, 2022.

\bibitem{10504275}
D.~Gürgünoğlu, E.~Björnson, and G.~Fodor, ``{Combating Inter-Operator Pilot Contamination in Reconfigurable Intelligent Surfaces Assisted Multi-Operator Networks},'' \emph{IEEE Trans. on Comm.}, pp. 1--1, 2024.

\bibitem{de2024beyond}
A.~Sousa~de Sena, M.~Rasti, N.~Huda~Mahmood, and M.~Latva-aho, ``{Beyond Diagonal RIS for Multi-Band Multi-Cell MIMO Networks: A Practical Frequency-Dependent Model and Performance Analysis},'' \emph{IEEE Trans. on Wireless Comm.}, vol.~24, no.~1, pp. 749--766, 2025.

\bibitem{9776539}
J.~Ye, A.~Kammoun, and M.-S. Alouini, ``{Reconfigurable Intelligent Surface Enabled Interference Nulling and Signal Power Maximization in mmWave Bands},'' \emph{IEEE Transactions on Wireless Communications}, vol.~21, no.~11, pp. 9096--9113, 2022.

\bibitem{9146170}
L.~Zhang, Y.~Wang, W.~Tao, Z.~Jia, T.~Song, and C.~Pan, ``{Intelligent Reflecting Surface Aided MIMO Cognitive Radio Systems},'' \emph{IEEE Trans. on Vehicular Tech.}, vol.~69, no.~10, pp. 11\,445--11\,457, 2020.

\bibitem{alexandropoulos2023ris}
G.~C. Alexandropoulos, D.-T. Phan-Huy, K.~D. Katsanos, M.~Crozzoli, H.~Wymeersch, P.~Popovski, P.~Ratajczak, Y.~B{\'e}n{\'e}dic, M.-H. Hamon, S.~H. Gonzalez, \emph{et~al.}, ``{RIS-enabled smart wireless environments: Deployment scenarios, network architecture, bandwidth and area of influence},'' \emph{EURASIP Journal on Wireless Communications and Networking}, vol. 2023, no.~1, p. 103, 2023.

\bibitem{shangguan2022design}
Q.~Shangguan, Z.~Chen, H.~Yang, S.~Cheng, W.~Yang, Z.~Yi, X.~Wu, S.~Wang, Y.~Yi, and P.~Wu, ``{Design of ultra-narrow band graphene refractive index sensor},'' \emph{Sensors}, vol.~22, no.~17, p. 6483, 2022.

\bibitem{10130575}
Z.~Li, Z.~Gao, and T.~Li, ``{Sensing User's Channel and Location With Terahertz Extra-Large Reconfigurable Intelligent Surface Under Hybrid-Field Beam Squint Effect},'' \emph{IEEE Journal of Selected Topics in Signal Processing}, vol.~17, no.~4, pp. 893--911, 2023.

\bibitem{9771341}
S.~H. Park, B.~Kim, D.~K. Kim, L.~Dai, K.-K. Wong, and C.-B. Chae, ``{Beam Squint in Ultra-Wideband mmWave Systems: RF Lens Array vs. Phase-Shifter-Based Array},'' \emph{IEEE Wireless Communications}, vol.~30, no.~4, pp. 82--89, 2023.

\bibitem{9321220}
D.~Dardari and D.~Massari, ``{Using MetaPrisms for Performance Improvement in Wireless Communications},'' \emph{IEEE Transactions on Wireless Communications}, vol.~20, no.~5, pp. 3295--3307, 2021.

\bibitem{9759366}
W.~Cai, R.~Liu, M.~Li, Y.~Liu, Q.~Wu, and Q.~Liu, ``{IRS-Assisted Multicell Multiband Systems: Practical Reflection Model and Joint Beamforming Design},'' \emph{{IEEE Transactions on Communications}}, vol.~70, no.~6, pp. 3897--3911, 2022.

\end{thebibliography}

\begin{biography}[{\includegraphics[width=1in,height=1.25in,clip,keepaspectratio]{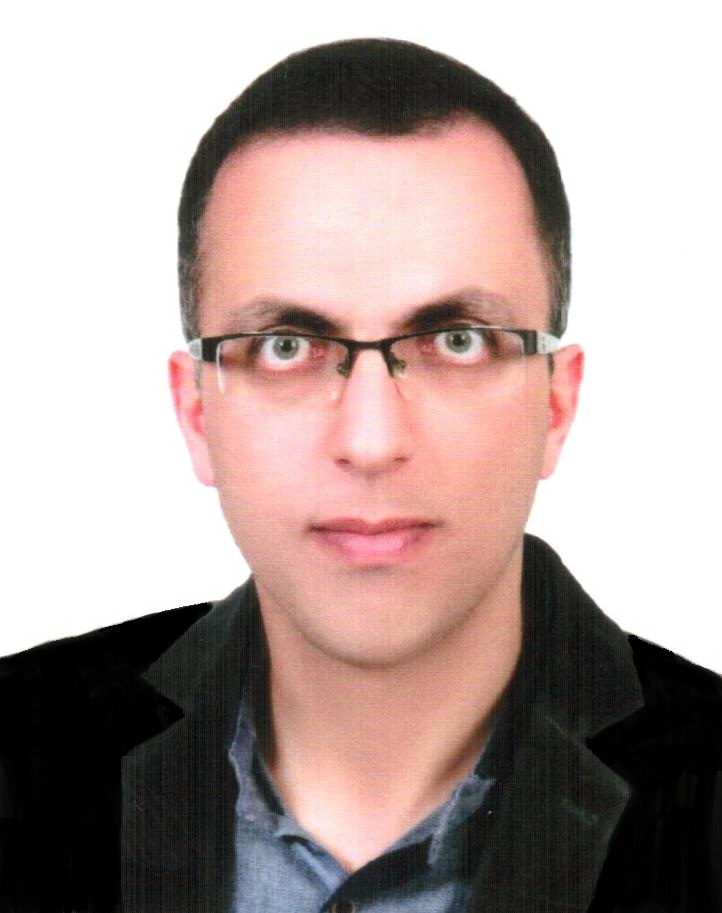}}]
		{Mehdi Monemi}
        (Member, IEEE) received the Ph.D. degree in electrical and computer engineering from Shiraz University, Shiraz, Iran, in 2014. He was a visiting researcher in the Department of Electrical and Computer Engineering, York University, Toronto, Canada from June 2019 to September 2019. He is currently a Postdoc researcher with the Centre for Wireless Communications (CWC), University of Oulu, Finland. His current research interests include resource allocation in 5G/6G networks, as well as the employment of ML in wireless networks.
\end{biography}

\vspace{-20pt}

\begin{biography}[{\includegraphics[width=1in,height=1.25in,clip,keepaspectratio]{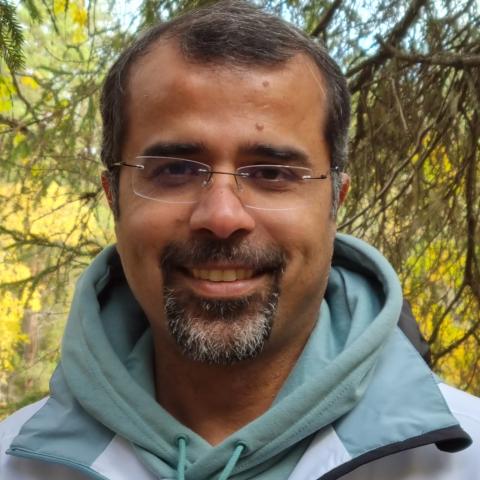}}]
		{Mehdi Rasti}
        (Senior Member, IEEE) received the Ph.D. degree from Tarbiat Modares University, Tehran, Iran, in 2009. He is currently an Associate Professor with the Centre for Wireless Communications, University of Oulu, Finland. %From 2012 to 2022, he was with the Department of Computer Engineering, Amirkabir University of Technology, Tehran, Iran.
        From February 2021 to January 2022, he was a Visiting Researcher with the Lappeenranta-Lahti University of Technology, Lappeenranta, Finland. His current research interests include radio resource allocation in IoT, Beyond 5G and 6G wireless networks.
\end{biography}

\vspace{-20pt}

\begin{biography}[{\includegraphics[width=1in,height=1.25in,clip,keepaspectratio]{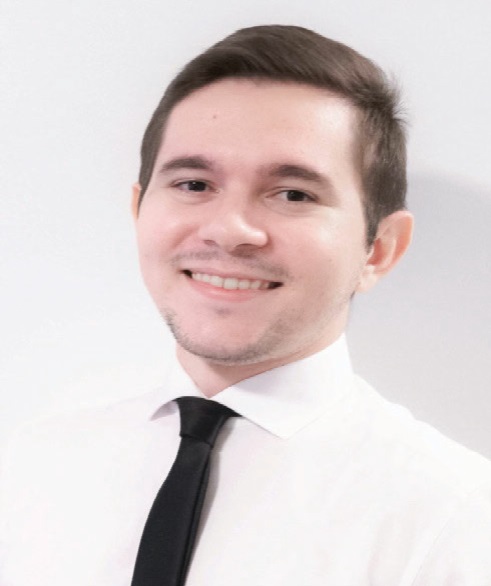}}]
		{Arthur Sousa de Sena}
        (Member, IEEE)  received
the D.Sc. degree (Hons.) in electrical engineering from LUT University, Finland, in 2022. Currently, he is a Postdoc Researcher with the Centre for Wireless Communication, University of Oulu, Finland, where he leads multiple research projects. %Previously, he was a Researcher with the AI and Digital Science Research Center, Technology Innovation Institute, Abu Dhabi, United Arab Emirates, from November 2022 to May 2023.
%From 2019 to 2022, he was a Junior Researcher with the Cyber-Physical Systems Group, LUT University. He has authored several peer-reviewed papers in prestigious journals and flagship  conferences. 
His research interests encompass topics including reconfigurable intelligent surfaces (RISs), next-generation multiple access schemes, integrated sensing and communication, and their intersection with
machine learning. %He received the Nokia Foundation Award in October 2020, the LUT Research Foundation Award in December 2020, and the IEEE Global Communications Conference (GLOBECOM) Best Paper Award in December 2022. 
He serves as an Editor for IEEE COMM. LETTERS.% and is a member of the IEEE Communications Society.
\end{biography}

\vspace{-20pt}

\begin{biography}[{\includegraphics[width=1in,height
=1.25in,clip,keepaspectratio]{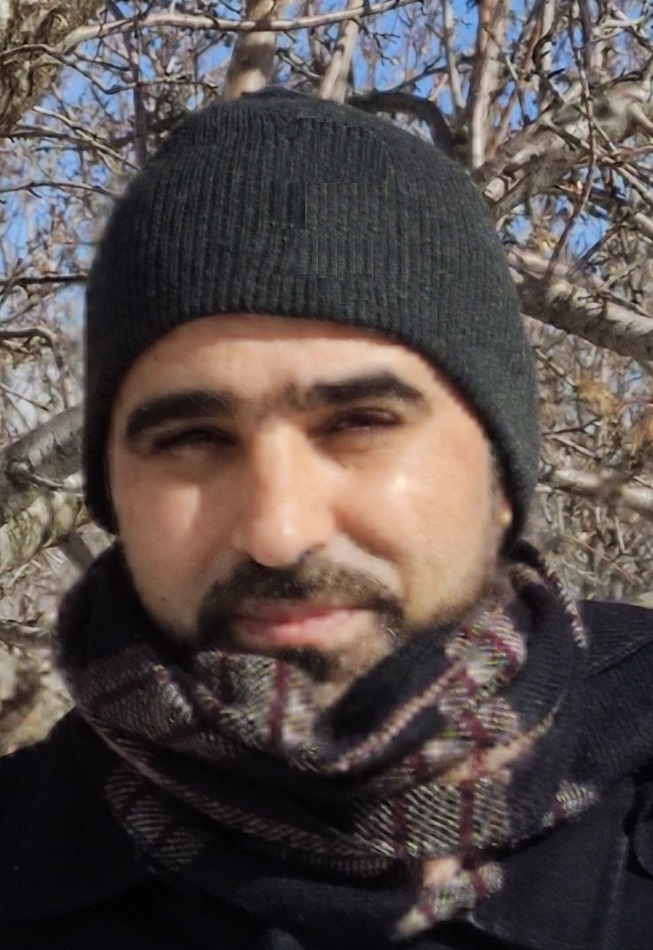}}]{Mohammad Amir Fallah}
		 received the Ph.D. degree in Electrical Engineering from Shiraz University, Shiraz, Iran, in 2013.
    He is currently an assistant professor with the Department of Engineering, Payame Noor University (PNU), Tehran, Iran. His current research interests include antenna and propagation, mobile computing, and the application of machine learning and artificial intelligence in wireless networks.
\end{biography}

\vspace{-20pt}

\begin{biography}[{\includegraphics[width=1in,height=1.25in,clip,keepaspectratio]{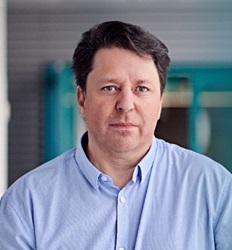}}]{Matti Latva-aho} (Fellow, IEEE) received Dr.Tech. (Hons.) degree in Electrical Engineering from the University of Oulu, Finland, in 1998. Prof. Latva-aho served as the Director of the Centre for Wireless Communications (CWC) from 1998 to 2006 and later as Head of the Department of Communication Engineering until August 2014. Currently, he is the Director of the National 6G Flagship Programme and Global Fellow at The University of Tokyo. He has published over 600 conference and journal publications.
\end{biography}

\vspace{-20pt}

\begin{biography}[{\includegraphics[width=1in,height=1.25in,clip,keepaspectratio]{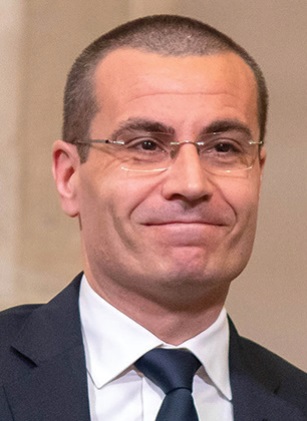}}]{Marco Di Renzo } (Fellow, IEEE) is a CNRS Research Director (Professor) and the Head of the Intelligent Physical Communications group with the Laboratory of Signals and Systems (L2S) at CNRS \& CentraleSupélec, Paris-Saclay University, Paris, France, as well as a Chair Professor in Telecommunications Engineering with the Centre for Telecommunications Research -- Department of Engineering, King’s College London, London, United Kingdom. In 2024, he was a France-Nokia Chair of Excellence in ICT at the University of Oulu, Finland. He served as the Editor-in-Chief of IEEE Communications Letters from 2019 to 2023, and he is currently serving as the Director of Journals of the IEEE Communications Society. %is a CNRS Research Director (Professor) and the Head of the Intelligent Physical Communications group in the Laboratory of Signals and Systems at Paris-Saclay University — CNRS and CentraleSupelec, Paris, France. He holds the 2023 France-Nokia Chair of Excellence in ICT (Finland), the 2023 Tan Chin Tuan Exchange Fellowship in Engineering at Nanyang Technological University (Singapore), and is the 2024 Ambassador of the European Association on Antennas and Propagation. He served as the Editor-in-Chief of IEEE Comm. Letters in 2019–2023, and he is currently serving as the Director of Journals of the IEEE Communications Society.
\end{biography}

\end{document}